\documentclass[aps,reprint,superscriptaddress,showpacs]{revtex4-1}

\usepackage{amsmath}
\usepackage{amssymb}
\usepackage{amsfonts}

\usepackage{slashed}
\usepackage{mathrsfs}
\usepackage[breaklinks]{hyperref}
\usepackage{graphicx}
\usepackage{subfigure}
\usepackage{multirow}

\begin{document}
%
%
\title{$CP$ violation in $B^\pm\to\pi^\pm\pi^+\pi^-$ in the region with low invariant mass of one $\pi^+\pi^-$ pair}
%
%
\author{Zhen-Hua~Zhang}
 \email[Email: ]{zhangzh@iopp.ccnu.edu.cn}
  \affiliation{Institute of Particle Physics, Huazhong Normal University, Wuhan 430079, China}
\author{Xin-Heng~Guo}
\email[Email: ]{xhguo@bnu.edu.cn}
\affiliation{College of Nuclear Science and Technology, Beijing Normal University, Beijing 100875, China}%
\author{Ya-Dong~Yang}
 \email[Corresponding author, Email: ]{yangyd@iopp.ccnu.edu.cn}
  \affiliation{Institute of Particle Physics, Huazhong Normal University, Wuhan 430079, China}
 \date{\today}
%
\begin{abstract}
Recently, the large $CP$ asymmetries in $B^\pm\to\pi^\pm\pi^+\pi^-$ decays were found by the LHCb Collaboration to localize in the region $m_{\pi^+\pi^-}^2<0.4~\text{GeV}^2$.
We find such large localized $CP$ asymmetries may be due to the interference between a light scalar and $\rho^0(770)$ intermediate resonances.
Consequently, we argue that the distribution of $CP$ asymmetries in the Dalitz plots of three-body $B$ decays could be very helpful for identifying the presence of the scalar resonance.
\end{abstract}
%
\pacs{11.30.Er, 13.25.Hw, 14.40.Nd}
%
\maketitle
%
%

Recently, the LHCb Collaboration found clear evidence for direct $CP$ violation in some three-body decay channels of $B$ mesons such as $B^\pm\to\pi^+\pi^-\pi^\pm$ and $B^\pm\to\pi^+\pi^- K^\pm$ \cite{deMiranda:2013kg,LHCb-CONF-2012-028}.
Intriguingly, large direct $CP$ asymmetries wrere found in some localized phase spaces of the two decay channels.
For $B^\pm\to\pi^+\pi^-\pi^\pm$, the $CP$ asymmetry in the region  $m^2_{\pi^+\pi^-~\text{low}}<0.4~\text{GeV}^2$ and $m^2_{\pi^+\pi^-~\text{high}}>15~\text{GeV}^2$
 is 
\footnote{For the decay channel $B^-\to \pi^-\pi^+\pi^-$, there are two identical  pions with negative charge.
When combining the momentum of each $\pi^-$ meson with that of the $\pi^+$ meson, we will have two Lorentz invariant mass squares which are usually different in values and are denoted by $m_{\pi^+\pi^-~\text{low}}^2$ and $m_{\pi^+\pi^-~\text{high}}^2$ in Ref. \cite{deMiranda:2013kg}, respectively.
Throughout this paper, we will denote $m_{\pi^+\pi^-~\text{low(high)}}^2$ as $s_{L(H)}$ for simplicity.}
\begin{equation}
 A_{CP}=+0.622\pm0.075\pm0.032\pm0.007,
\end{equation}
 while in the region $m^2_{\pi^+\pi^-~\text{low}}<0.4~\text{GeV}^2$ and $m^2_{\pi^+\pi^-~\text{high}}<15~\text{GeV}^2$, no large $CP$ asymmetry was observed 
 \footnote{In fact, a large $CP$ asymmetry difference in upper and lower parts of localized region $m_{\pi^+\pi^-}^2<0.4~\text{GeV}^2$ was also observed in $B^\pm\to\pi^+\pi^- K^\pm$.}.

In this paper, we will show that the localized large $CP$ asymmetry may arise from the interference between intermediate $\rho^0$ and another scalar meson nearby in the three-body decays.

%
%

It is known that the scalar resonance is very difficult to identify because of its large width.
In the following, we will show that the localized $CP$ asymmetries could be very helpful for identifying a scalar resonance which interferes with the vector one nearby.
We will consider a $B$ meson weak decay process, $B\to M_1M_2 M_3$, where $M_i$ ($i=1, 2, 3$) is a light pseudoscalar meson.
If this process is dominated by a resonance $X$ in a certain region of its Dalitz plot, then it will be very difficult to tell whether another resonance exists close to $X$.
We assume that $X$ is a vector meson, the possible resonance $Y$ nearby is a scalar meson, and both $X$ and $Y$ decay to $M_1M_2$.
The amplitude for $B\to M_1M_2 M_3$ around the $Y$ resonance region can be expressed as
\begin{equation}
\label{GeneralAmplitude}
\mathcal{M}=\mathcal{M}_X+\mathcal{M}_Y\text{e}^{i\delta},
\end{equation}
where $\delta$ is a relative strong phase, $\mathcal{M}_X$ and $\mathcal{M}_Y$ are the amplitudes for $B\to X M_3\to M_1M_2 M_3$ and $B\to Y M_3\to M_1M_2 M_3$, respectively, and they take the form
\begin{eqnarray}
\label{MX}
\mathcal{M}_X&=&\frac{g_X}{s_X}(s_{13}-\hat{s}_{13})(\hat{T}_{X}+\hat{P}_{X}\text{e}^{i\delta_X}\text{e}^{i\phi}),\\
\label{MY}
\mathcal{M}_Y&=&\frac{g_Y}{s_Y}(T_{Y}+P_{Y}\text{e}^{i\delta_Y}\text{e}^{i\phi}).
\end{eqnarray}
In the above two equations, $s_{ij}$ ($i, j=1, 2, 3$) is the invariant mass squared of mesons $M_i$ and $M_j$, $g_{X(Y)}$ is the effective coupling constant for the strong decay $X(Y)\to M_1 M_2$, $s_{X(Y)}$ is the reciprocal of the propagator of $X$ ($Y$) which takes the form $s_{12}-m_{X(Y)}^2+i\sqrt{s_{12}}\Gamma_{X(Y)}(s_{12})$ 
\footnote{The $s$-dependent decay width of $Z$ ($Z$ can be $X$ or $Y$) takes a general form
\begin{equation*}
\Gamma_{Z}(s)=\left[\frac{g_{Z}(s)}{g_{Z}}\right]^2\!\!\left[\frac{\lambda(\sqrt{s})/s}{\lambda(m_{Z})/m_{Z}^2}\right]^{\xi_{Z}}\!\!\left[\frac{m_{Z}^2}{s}\right]\!\Gamma_{Z},
\end{equation*}
where $\xi_{Z}=S_{Z}+1/2$, with $S_{Z}$ being the spin of $Z$, $\lambda(x)=[x^2\!-\!(m_{M_1}\!+\!m_{M_2})^2][x^2-(m_{M_1}\!-\!m_{M_2})^2]$.
In the numerical calculation of this paper, we simply set $g_{Z}(s)=g_{Z}$.
}, 
$T_{X(Y)}$ and $P_{X(Y)}$ are the tree and the penguin amplitudes for the decay $B\to X(Y) M_3$, $\hat{T}_X=T_X/(\varepsilon^\ast\cdot p_B)$ and $\hat{P}_X=P_X/(\varepsilon^\ast\cdot p_B)$ with $\varepsilon$ being the polarization vector of the meson $X$
\footnote{
Generally, both the tree and penguin amplitudes of $B\to XM_3$ should be proportional to $\varepsilon^\ast\cdot p_B$. 
When considering a process in which the vector meson $X$ predominantly decays into $M_1M_2$, one should replace the polarization vector $\varepsilon^\ast_\mu$ by 
\begin{equation*}
\frac{g_X}{s_X}(p_{M_1}\!\!-\!p_{M_2})^\nu\!\!\left[g_{\mu\nu}\!-\!\frac{(p_{M_1}+p_{M_2})_{\mu}(p_{M_1}+p_{M_2})_{\nu}}{s_{12}}\right].
\end{equation*} 
Taking the dot product of this term and $p_B$, one will get $g_X(s_{13}-\hat{s}_{13})/s_X$.
}, 
 $\delta_{X,Y}$ are the relative strong phases between the tree and the penguin amplitudes, $\phi$ is the weak phase, and $\hat{s}_{13}$ is the midpoint of the allowed range of $s_{13}$, i.e., $\hat{s}_{13}=(s_{13,\text{max}}+s_{13,\text{min}})/2$, with $s_{13,\text{max}}$ and $s_{13,\text{min}}$ being the maximum and minimum values of $s_{13}$  for fixed $s_{12}$.
 One can check that
\begin{equation}
\label{eq:s13}
 \hat{s}_{13}\!\!=\!\!\frac{1}{2}\!\!\left[\!\Big(\!m_B^2\!+\!\!\sum_{i}\!m_{M_i}^2\!\!-s_{12}\!\Big)\!\!+\!\!\frac{(m_{M_1}^2\!\!\!-\!m_{M_2}^2)\!(m_B^2\!\!-\!m_{M_3}^2)}{s_{12}}\!\right].
\end{equation}
 The second term in Eq. (\ref{eq:s13}) is small compared with the first one, since usually $(m_{M_1}^2-m_{M_2}^2)\ll m_X^2$.
Throughout this paper, we will denote the momentum, the mass, and the decay width of a particle $X$ by $p_X$, $m_X$, and $\Gamma_X$, respectively. 

As aforementioned, we will focus on the region around the $Y$ resonance, i.e., $m_Y-\Delta_1<\sqrt{s_{12}}<m_Y+\Delta_2$, where $\Delta_1$ and $\Delta_2$ are of the order of $\Gamma_Y$.
We also require that $m_Y-\Delta_1 > m_X+\Gamma_X$ (if $m_Y>m_X$) or $m_Y+\Delta_2 < m_X-\Gamma_X$ (if $m_Y<m_X$), so that these two resonances have competitive contributions in this region. 

For the region of phase space (denoted by $\omega$) where the two amplitudes $\mathcal{M}_X$ and $\mathcal{M}_Y$ are competitive, the direct $CP$ violation parameter is found to be
\begin{equation}
\label{ACPomega}
A_{CP}^{\omega}=\frac{\mathcal{S}_-^\omega+\mathcal{A}_-^\omega}{\mathcal{S}_+^\omega+\mathcal{A}_+^\omega},
\end{equation}
where
\begin{widetext}
\begin{eqnarray}
\mathcal{S}_-^\omega&=&-2\sin\phi\!\!\int_{\omega}\!\!\text{d}s_{12}\text{d}s_{13}\left[\tilde{T}_X\tilde{P}_X\sin\delta_X+\tilde{T}_Y\tilde{P}_Y\sin\delta_Y\right],\\
\mathcal{S}_+^\omega&=&\!\!\int_{\omega}\!\!\text{d}s_{12}\text{d}s_{13}\left[\tilde{T}_X^2+\tilde{T}_Y^2+\tilde{P}_X^2+\tilde{P}_Y^2+2\cos\phi\left(\tilde{T}_X\tilde{P}_X\cos\delta_X+\tilde{T}_Y\tilde{P}_Y\cos\delta_Y\right)\right],\\
\mathcal{A}_-^\omega&=&-2\sin\phi\!\!\int_{\omega}\!\!\text{d}s_{12}\text{d}s_{13}\left[\tilde{T}_X\tilde{P}_Y\sin(\delta_Y+\tilde{\delta})+\tilde{T}_Y\tilde{P}_X\sin(\delta_X-\tilde{\delta})\right],\\
\mathcal{A}_+^\omega&=&2\!\!\int_{\omega}\!\!\text{d}s_{12}\text{d}s_{13}\Big\{\tilde{T}_X\tilde{T}_Y\cos\tilde{\delta}+\tilde{P}_X\tilde{P}_Y\cos(\delta_X-\delta_Y-\tilde{\delta})+\cos\phi\big[\tilde{T}_X\tilde{P}_Y\cos(\delta_Y+\tilde{\delta})+\tilde{T}_Y\tilde{P}_X\cos(\tilde{\delta}-\delta_X)\big]\Big\},
\end{eqnarray}
\end{widetext}
with $\tilde{\delta}=\delta+\arg (s_X)-\arg (s_Y)$, and
\begin{eqnarray}
\label{TX}
\tilde{T}_X&=&\frac{g_X}{|s_X|}(s_{13}-\hat{s}_{13})\hat{T}_{X},\\
\label{TY}
\tilde{T}_Y&=&\frac{g_Y}{|s_Y|}T_{Y},
\end{eqnarray}
and similar definitions for $\tilde{P}_X$ and $\tilde{P}_Y$.

From Eqs. (\ref{MX}) and (\ref{MY}), one can easily check the following relations,
\begin{eqnarray}
\label{MXproperty}
 \mathcal{M}_X(s_{13})&=&-\mathcal{M}_X(\bar{s}_{13}),\\
\label{MYproperty}
 \mathcal{M}_Y(s_{13})&=&\mathcal{M}_Y(\bar{s}_{13}),
\end{eqnarray}
where $\bar{s}_{13}=2\hat{s}_{13}-s_{13}$.
These relations allow us to divide naturally the region around the $Y$ resonance into two parts: $\Omega$ and $\bar{\Omega}$, where $\Omega$ is for $s_{13}>\hat{s}_{13}$ and $\bar{\Omega}$ is for $s_{13}<\hat{s}_{13}$.
From Eqs. (\ref{MXproperty}) and (\ref{MYproperty}), we can derive the following relations between $\Omega$ and $\bar{\Omega}$ phase spaces:
\begin{eqnarray}
\label{Ssymmetry}
&&\mathcal{S}_\pm^\Omega=\mathcal{S}_\pm^{\bar{\Omega}}, \\
\label{Asymmetry}
&&\mathcal{A}_\pm^\Omega=-\mathcal{A}_\pm^{\bar{\Omega}}.
\end{eqnarray}
Besides the $CP$ violation in Eq. (\ref{ACPomega}), we define four other quantities 
\begin{eqnarray}
\label{Rpm}
R_{\pm}=\frac{(N^\Omega\pm \bar{N}^{\Omega})-(N^{\bar{\Omega}}\pm \bar{N}^{\bar{\Omega}})}{(N^\Omega\pm \bar{N}^{\Omega})+(N^{\bar{\Omega}}\pm \bar{N}^{\bar{\Omega}})},\\
\label{Wpm}
W_{\pm}=\frac{(N^\Omega - \bar{N}^{\Omega})\pm(N^{\bar{\Omega}}- \bar{N}^{\bar{\Omega}})}{(N^\Omega+ \bar{N}^{\Omega})\pm(N^{\bar{\Omega}}+ \bar{N}^{\bar{\Omega}})},
\end{eqnarray}
where all the $N$'s ($\bar{N}$'s) are the event numbers of $B\to M_1M_2 M_3$ ($\overline{B}\to\overline{M}_1\overline{M}_2\overline{M}_3$) in the corresponding phase space.
With Eqs. (\ref{Ssymmetry}) and (\ref{Asymmetry}), one can easily check
\begin{eqnarray}
R_\pm=\mathcal{A}_\pm^\Omega/\mathcal{S}_\pm^\Omega,\\
W_+=\mathcal{S}_-^\Omega/\mathcal{S}_+^\Omega,\\
W_-=\mathcal{A}_-^\Omega/\mathcal{A}_+^\Omega.
\end{eqnarray}
Note that $R_-$ is independent of the weak phase $\phi$ and $|R_+|<1$ by definition.
So far, we have six quantities: $A_{CP}^\Omega$, $A_{CP}^{\bar{\Omega}}$, $R_{\pm}$, and $W_{\pm}$, but only three of them are independent.
Alternatively, the $CP$ violations in phase spaces $\Omega$ and $\bar{\Omega}$ read
\begin{eqnarray}
\label{alterCP}
A_{CP}^{\Omega}=W_+\frac{1+R_-}{1+R_+},\\
\label{alterCPbar}
A_{CP}^{\bar{\Omega}}=W_+\frac{1-R_-}{1-R_+}.
\end{eqnarray}

One can see that the $CP$ asymmetries in these two regions can be very different because of the existence of the antisymmetric terms $\mathcal{A}_\pm^\Omega$ under the interchange of $\Omega$ and $\bar{\Omega}$. 
These antisymmetric terms exist because the two resonances $X$ and $Y$ have different spins.
If both $X$ and $Y$ have the same spin, then $A_\pm^\Omega\equiv A_\pm^{\bar{\Omega}}$, and one would observe that the $CP$ asymmetries in the two regions equal each other. 
One may argue that we cannot exclude the possibility that the $CP$ violations may be the same in phase spaces $\Omega$ and $\bar{\Omega}$ even if $X$ and $Y$ are vector and scalar mesons, respectively.
This is indeed true, and the $CP$ asymmetry difference between $\Omega$ and $\bar{\Omega}$ cannot be used as a probe of the scalar resonance in this situation.
However, both $R_-$ and $R_+$ become good probes.
The nonzero values of $R_-$ and $R_+$ will imply the presence of the scalar resonance $Y$.
One can check that if $Y$ is a vector resonance, then both $R_+$ and $R_-$ equal zero.

Furthermore, there is also an alternative criteria that can be used to identify the resonance of $Y$.
Since the amplitude $\mathcal{M}_X$ becomes very small when $s_{13}$ is close to $\hat{s}_{13}$, the amplitude $\mathcal{M}_Y$ will be dominant over $\mathcal{M}_X$, and then one should observe a larger density of events when $s_{12}\sim m_Y^2$ than when $s_{12}\sim m_X^2$, on the condition that $s_{13}$ is close to $\hat{s}_{13}$
\footnote{However, with this criteria alone, we cannot know the spin of $Y$. For example, if $Y$ is a tensor resonance, one can also observe a larger event density at $s_{12}\sim m_Y^2$ than that at $s_{12}\sim m_X^2$ when $s_{13}$ is close to $\hat{s}_{13}$}.

We have used the transverse approximation for the propagator of the vector meson $X$.
 The numerator of the propagator of $X$ is $g_{\mu\nu}-{k_\mu k_\nu}/{s_{12}}$ (up to a phase factor) with $k=p_{M_1}+p_{M_2}$. 
 This has a different off-shell behavior from the propagator for a  pointlike vector particle, $g_{\mu\nu}-{k_\mu k_\nu}/{m_X^2}$.
In fact, since hadrons are not pointlike particles, one inevitably confronts this kind of ambiguity when dealing with vector mesons.
If we instead use the latter form of the propagator for the vector resonance, we should add to $\hat{s}_{13}$ in Eq. (\ref{TX}) with a term 
\begin{equation*}
\frac{m_X^2-s_{12}}{2m_X^2}(m_{M_1}^2-m_{M_2}^2)(m_B^2-m_{M_3}^2-s_{12}).
\end{equation*}
 When $s_{13}$ is far away from $\hat{s}_{13}$, this term is small compared with $s_{13}-\hat{s}_{13}$.
It only becomes comparable with $s_{13}-\hat{s}_{13}$ when $s_{13}$ is close to $\hat{s}_{13}$. However, in this case, $\mathcal{M}_X$ is small compared with $\mathcal{M}_Y$.
Therefore, we are free to use the transverse approximation for the propagator of the vector meson.

We want to mention the following two special cases:

\textit{Case 1}: Both $\delta_X$ and $\delta_Y$ are very small, but $\delta$ is not small.
In this situation, both $S_-^\Omega$ and $A_-^\Omega$ are small and can be neglected safely.
One would observe that $A_{CP}^{\Omega}$ and $A_{CP}^{\bar{\Omega}}$ have opposite signs.

\textit{Case 2}: All the three strong phases $\delta_X$, $\delta_Y$, and $\delta$ are very small.
In this situation, the $CP$ violation parameters in both regions will be very close to zero.
Then, one cannot identify the presence of $Y$ just through the measurement of $CP$ violation parameters.
However, one can still identify the presence of $Y$ by measuring $R_+$.
The nonzero value of $R_+$ indicates the existence of $B\to Y M_3\to M_1 M_2 M_3$.

In the above discussion, we have assumed that $X$ and $Y$ are vector and scalar mesons, respectively.
One can arrive at a similar conclusion by reversing their spins.
Our analysis can also be generalized to situations when both $X$ and $Y$ have arbitrary spins.
If $X$ is a resonance with spin $J$, the corresponding amplitude $\mathcal{M}_X$ would be proportional to $(s_{13}-\hat{s}_{13}^{(1)})(s_{13}-\hat{s}_{13}^{(2)})\cdots(s_{13}-\hat{s}_{13}^{(J)})$, where $\hat{s}_{13}^{(1)}$, $\hat{s}_{13}^{(2)}$, \dots, $\hat{s}_{13}^{(J)}$ lie within the allowed range of $s_{13}$. 
Take $X$ as a tensor meson ($Y$ still a scalar meson) for example.
In this situation, $\mathcal{M}_X\propto(s_{13}-\hat{s}_{13}^{(1)})(s_{13}-\hat{s}_{13}^{(2)})$, where $\hat{s}_{13}^{(1)}=\hat{s}_{13}-\Delta_{13}/\sqrt{3}$ and $\hat{s}_{13}^{(2)}=\hat{s}_{13}+\Delta_{13}/\sqrt{3}$ with $\Delta_{13}=(s_{13,\text{max}}-s_{13,\text{min}})/2$.
One would observe that there is a large difference of $CP$ asymmetries between the middle part ($\hat{s}_{13}^{(1)}<s_{13}<\hat{s}_{13}^{(2)}$) and the other two parts ($s_{13}<\hat{s}_{13}^{(1)}$ and $s_{13}>\hat{s}_{13}^{(2)}$).

%
%
Now we are ready to show that the large localized CP asymmetries observed by LHCb in $B^\pm\to\pi^\pm\pi^+\pi^-$ can be interpreted as the interference of $\rho^0$ and $f_0(500)$. 
The LHCb Collaboration found that for $B^\pm\to\pi^\pm\pi^+\pi^-$, the dominant resonance is the vector meson $\rho^0(770)$ \cite{deMiranda:2013kg}.
In the region $s_L<0.4~\text{GeV}^2$, there is a large difference of $CP$ asymmetries between the upper ($s_H>15~\text{GeV}^2$) and the lower ($s_H<15~\text{GeV}^2$) parts.
In the following, we will denote these two parts by $\Omega'$ and $\bar{\Omega}'$, respectively.
Note that $15~\text{GeV}^2$ is very close to $\hat{s}_H$, which is about $14~\text{GeV}^2$ for $s_L<0.4~\text{GeV}^2$.
According to the above analysis, one immediately concludes that there is a resonance with spin 0 lying in the region $s_L<0.4~\text{GeV}^2$. 
From PDG \cite{Beringer:1900zz}, we know that this particle could be $f_0(500)$.
In the following, we will show that by including the amplitudes for $B^\pm\to f_0(500)\pi^\pm\to\pi^+\pi^-\pi^\pm$ and $B^\pm\to\rho^0\pi^\pm\to\pi^+\pi^-\pi^\pm$, the observed $CP$ violation behavior can be understood.

We assume that the two amplitudes of $B^\pm\to f_0(500)\pi^\pm\to\pi^+\pi^-\pi^\pm$ and $B^\pm\to\rho^0\pi^\pm\to\pi^+\pi^-\pi^\pm$ are dominant for $s_L<0.4~\text{GeV}^2$.
They can be expressed as
\begin{eqnarray}
\mathcal{M}_{\rho^0}
&=&\frac{g_{\rho\pi\pi}}{s_{\rho^0}}(s_H-\hat{s}_H)\frac{\mathcal{M}_{B^-\to\pi^- \rho}}{\varepsilon^\ast\cdot p_B},\\
\mathcal{M}_{f_0}
&=&\frac{g_{f_0\pi\pi}}{s_{f_0}}\mathcal{M}_{B^-\to\pi^-f_0},
\end{eqnarray}
where $f_0$ represents $f_0(500)$ and $\mathcal{M}_{B^-\to\pi^- \rho}$ and $\mathcal{M}_{B^-\to\pi^-f_0}$ are the amplitudes for $B^-\to\pi^- \rho$ and $B^-\to\pi^-f_0$, respectively.

%
\begin{figure}
\includegraphics[width=0.45\textwidth]{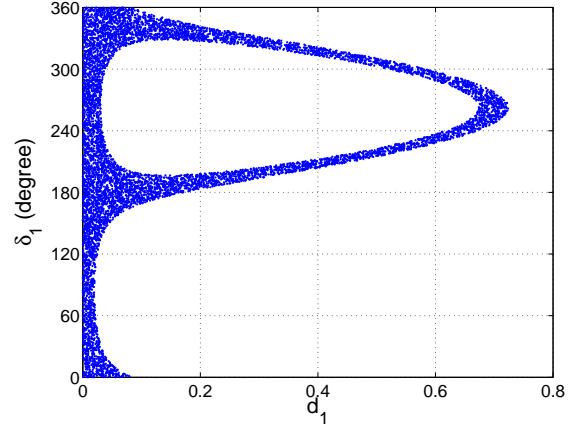}
\caption{\label{d1delta1}Allowed region for $d_1$ and $\delta_1$. If plotted in polar coordinate system, one can find the allowed region is actually a circular ring crossing the origin.}
\end{figure}
%

With the effective Hamiltonian for the weak transition $b\to q\bar{q}d$ \cite{Buchalla:1995vs},  one can obtain the decay amplitudes for $B\to\rho\pi$ and $B\to f_0(500)\pi$, which can be expressed as (a common factor $G_F/\sqrt{2}$ has been neglected)
\begin{widetext}
\begin{eqnarray}
\mathcal{M}_{B^-\to\pi^-\rho^0}&=&V_{ub}V_{ud}^{\ast}[a_2X^{(B^-\rho^0,\pi^-)}+a_1X_u^{(B^-\pi^-,\rho^0)}]-V_{tb}V_{td}^{\ast}\bigg\{\left[-a_4+\frac{3}{2}a_7+\frac{3}{2}a_9+\frac{1}{2}a_{10}\right]X_{u}^{(B^-\pi^-,\rho^0)}\nonumber\\
&&+\left[a_4+a_{10}+\left(\frac{m_B}{m_\pi}d_1\text{e}^{i\delta_1}-2\right)\frac{(a_6+a_8)m_\pi^2}{(m_d+m_u)(m_b+m_u)}\right]X^{(B^-\rho^0,\pi^-)}\bigg\},\\
\mathcal{M}_{B^-\to f_0\pi^-}
&=&V_{ub}V_{ud}^\ast a_2 X^{(B^-f_0,\pi^-)}-V_{tb}V_{td}^\ast\left\{
\left[a_4+a_{10}+\left(\frac{m_B}{m_\pi}d_2\text{e}^{i\delta_2}-2\right)\frac{(a_6+a_8)m_\pi^2}{(m_u+m_d)(m_b+m_u)}\right]X^{(B^-f_0,\pi^-)}\right\},\nonumber\\
\end{eqnarray}
\end{widetext}
where all the $a_i$'s are built up from the effective Wilson coefficients $C'_{i}$s, and take the form $a_i=C'_i+C'_{i+1}/N_c$ for odd $i$ and $a_i=C'_i+C'_{i-1}/N_c$ for even $i$, the notation $X$ for matrix elements is borrowed from Ref. \cite{Chen:1999nxa}.
For example, $X^{(B^-\rho^0,\pi^-)}$ is defined as $\langle \pi^-|(\bar{d}u)_{V-A}|0\rangle\langle \rho^0|(\bar{u}b)_{V-A}|B^-\rangle$.
These matrix elements can be parametrized as the products of decay constants and form factors.
For numerical results, we use $F_1^{B\to\pi}(0)=0.25$ and $A_0^{B\to\rho}(0)=0.28$ \cite{Cheng:2003sm}.
We also simply set $F^{B\to f_0}(0)=0.3$.

Terms containing $d_1$ and $d_2$ come from annihilation terms, which are proportional to  $X^{(B^-,\rho^0\pi^-)}$ or $X^{(B^-,f_0\pi^-)}$.
Usually, annihilation terms are suppressed by at least a factor $\Lambda_{QCD}/m_b$, so that one can neglect them safely.
However, there are also annihilation terms that are enhanced by a chiral factor, $m_B^2/[(m_b+m_u)(m_d+m_u)]$.
This kind of term should be taken into account with proper parametrization.
According to our parametrization, $d_1$ and $d_2$ should be, at most, order one.
Because of multiple soft scattering, annihilation diagrams may also give rise to strong phases.
This explains the appearance of $\delta_1$ and $\delta_2$.
For the effective Wilson coefficients, we will adopt the set of coefficients in Ref. \cite{Cheng:1998uy}.

We have the following five free parameters: $\delta$, $d_1$, $\delta_1$, $d_2$, and $\delta_2$.
Since $d_1$ and $\delta_1$ are related to the chiral enhancement, this makes them potentially sensitive to the branching ratio of $B^+\to\rho^0\pi^+$.
Thus,  these two parameters can be constrained by the experimental data for the branching ratio of $B^+\to\rho^0\pi^+$.
We use the following experimental data to determine the allowed region for $d_1$ and $\delta_1$ \cite{Beringer:1900zz}:
\begin{equation}
\mathcal{BR}(B^+\to\rho^0\pi^+)=(8.3\pm1.2)\times10^{-6}.
\end{equation}
The results are shown in FIG. \ref{d1delta1}.

For given allowed values of $d_1$ and $\delta_1$, we should determine the allowed regions for the other three parameters with the aid of the data,
\begin{eqnarray}
A_{CP}^{\Omega'}=+0.62\pm0.10,\\
A_{CP}^{\bar{\Omega}'}=-0.05\pm0.05.
\end{eqnarray}
In Table \ref{AllowedRegion}, we show the allowed regions of $\delta$, $d_2$, and $\delta_2$ for given values of $d_1$ and $\delta_1$.
Note that the allowed regions of these three parameters are in fact correlated.
What we show in the table are actually the largest ranges.  
The correlated allowed region of these parameters is a subset of the direct combined region shown in Table \ref{AllowedRegion}.

The change of input parameters may change the allowed regions of the parameters shown in Table \ref{AllowedRegion}, but it does not change the conclusion that the large $CP$ asymmetry difference between phase spaces $\Omega'$ and $\bar{\Omega}'$ is caused by the interference of $\rho^0$ and $f_0(500)$.
We also anticipate that $R_\pm$ should be nonzero, and this can be checked by the data very easily.
Because $A_{CP}^{\bar{\Omega}'}$ (so that $A_{CP}^{\bar{\Omega}}$) is very small, we also predict that $R_-$ is a little bit larger than 1.

We confronted two resonances during our calculations, $\rho^0(770)$ and $f_0(500)$. 
The masses and total decay widths of these two resonances in our numerical calculation are (in GeV) \cite{Beringer:1900zz}
\begin{eqnarray}
&&m_{\rho^0(770)}=0.775, ~~~~\Gamma_{\rho^0(770)}=0.149,\nonumber\\
&&m_{f_0(500)}=0.500, ~~~~\Gamma_{f_0(500)}=0.500.\nonumber
\end{eqnarray}
Since the nature of $f_0(500)$ is not known yet, our numerical calculation here should be regarded as an estimation.
We also used the factorization hypothesis during our calculations of amplitudes corresponding to the two intermediate resonances, $\rho^0$ and $f_0(500)$. 
Since the $\rho^0$ meson is not on the mass shell,  the calculation with this hypothesis is clearly not accurate.
However, since the interested area of the phase space is not far away form the $\rho^0$ mass shell, the  factorization hypothesis is still good enough for an estimation.

In summary, we have shown that the interference of two intermediate resonances with different spins can result in a $CP$ violation difference in the corresponding phase space, which can be used as a method to identify the scalar resonance that is close to a vector one. 
With this method, we show that the recently observed large $CP$ asymmetry difference in $B^\pm\to\pi^\pm\pi^+\pi^-$ decays localized in the region $m_{\pi^+\pi^-}<0.4\text{GeV}^2$ indicates the existence of a scalar resonance, which can be identified as $f_0(500)$.

%
\begin{table}[htdp]
\caption{Allowed regions of $\delta$, $\delta_2$, and $d_2$ with given values of $d_1$ and $\delta_1$.\label{AllowedRegion}}
\begin{center}
\begin{tabular}{c c c c}
\hline
$(d_1, \delta_1)$ & $\delta$ & $\delta_2$ & $d_2$ \\
\hline
$(0.7,260^\circ)$ & $(3^\circ, 178^\circ)$ & $(-2^\circ,36^\circ)\cup(114^\circ,153^\circ)$& $(0.2,0.6)$\\
\hline
$(0.2,190^\circ)$ & $(-31^\circ, 54^\circ)$ & $(-1^\circ, 26^\circ)$ & $(0.1, 0.4)$\\
\hline
$(0.2,330^\circ)$ & $(141^\circ, 209^\circ)$ & $(133^\circ,153^\circ)$ & $(0.1,0.4)$\\
\hline
\end{tabular}
\end{center}
\label{default}
\end{table}%
%

%
\begin{acknowledgments}
This work was partially supported by National Natural Science Foundation of China under Contract No. 10975018, No. 11175020, No. 11225523, No. 11275025 and the Fundamental Research Funds for the Central Universities in China.
One of us (Z.H.Z.) also thanks the hospitality of Professor Xin-Nian Wang at Huazhong Normal University.
\end{acknowledgments}
%

\bibliography{zzh}

\end{document}